# Convergence of the discrete dipole approximation.
# II. An extrapolation technique to increase the accuracy.

**Maxim A. Yurkin**

Faculty of Science, Section Computational Science, of the University of Amsterdam,
Kruislaan 403, 1098 SJ, Amsterdam, The Netherlands
and
Institute of Chemical Kinetics and Combustion, Siberian Branch of the Russian Academy of
Sciences, Institutskaya 3, Novosibirsk 630090 Russia
myurkin@science.uva.nl

**Valeri P. Maltsev**

Institute of Chemical Kinetics and Combustion, Siberian Branch of the Russian Academy of
Sciences, Institutskaya 3, Novosibirsk 630090 Russia
and
Novosibirsk State University, Pirogova Str. 2, 630090, Novosibirsk, Russia

**Alfons G. Hoekstra**

Faculty of Science, Section Computational Science, of the University of Amsterdam,
Kruislaan 403, 1098 SJ, Amsterdam, The Netherlands
alfons@sciene.uva.nl

## Abstract

We propose an extrapolation technique that allows accuracy improvement of the discrete dipole approximation computations. The performance of this technique was studied empirically based on extensive simulations for 5 test cases using many different discretizations. The quality of the extrapolation improves with refining discretization reaching extraordinary performance especially for cubically shaped particles. A two order of magnitude decrease of error was demonstrated. We also propose estimates of the extrapolation error, which were proven to be reliable. Finally we propose a simple method to directly separate shape and discretization errors and illustrated this for one test case.

# 1. Introduction

The discrete dipole approximation (DDA) is a well-known method to solve the light scattering problem for arbitrary shaped particles. Since its introduction by Purcell and Pennypacker[1] it has been improved constantly. The formulation of DDA summarized by Draine and Flatau[2] more than 10 years ago is still most widely used for different applications,[3] partly due to the publicly available high-quality and user-friendly code DDSCAT.[4]

DDA directly discretizes the volume of the scatterer and hence is applicable to arbitrary shaped particles. However, the drawback of this discretization is the extreme computational complexity of DDA of $O(N^2)$, where $N$ is the number of dipoles. This complexity is decreased to $O(N\log N)$ by advanced numerical techniques.[2,5] Still the usual application strategy for DDA is "single computation", where a discretization is chosen based on available computational resources and some empirical estimates of the expected errors.[3,4] These error estimates are based on a limited number of benchmark calculations[3] and hence are external to the light scattering problem under investigation. Such error estimates have evident drawbacks, however no better alternative is available.

Usually errors in DDA are studied as a function of the size parameter of the scatterer $x$ (at a constant or few different values of $N$), e.g. [2,6]. Only several papers directly present errors versus discretization parameter (e.g. $d$ – the size of a single dipole).[7-15] The range of $d$ typically studied in those papers is limited to a 5 times difference between minimum and maximum values (with the exception of two papers[9,10] where it is 15 times). Only two papers[7,15] use extrapolation (to zero $d$) to get an exact result of some measured quantity, however they use the simplest linear extrapolation without any theoretical foundation nor discussion of its capabilities.

It is acknowledged for a long time that DDA errors are due to two different factors: *shape* (it is not always possible to describe the particle shape exactly by a collection of cubical cells) and *discretization* (finite size of each cell).[6] However, the question which of them is more important in different cases is still open. A discussion on this issue spanned through several papers[16-20] that have not reached any definite conclusions yet. The uncertainty is due to the indirect methods used that have inherent interpretation problems.

In accompanying paper,[21] that from now on we will refer to as Paper 1, we performed a theoretical analysis of DDA convergence when refining the discretization. It provides the basis for this paper, where an extrapolation technique is introduced (Section 2) to improve the accuracy of DDA computations. We thoroughly discuss all free parameters that influence extrapolation performance and provide a step-by-step prescription, which can be used with any existing DDA code without any modifications. It is important to note that although Paper 1 provides a firm theoretical background, it is not necessary to go through all theoretical details to understand and apply the extrapolation technique that we introduce here. In Section 3 we present extensive numerical results of DDA computations for 5 different scatterers using many different discretizations. These results are discussed in Section 4 to evaluate the performance of the extrapolation technique. We also propose a new method to directly separate shape and discretization errors of DDA (described and illustrated in Section 3.B). The results and possible applications are discussed in Section 4. We formulate the conclusions of the paper in Section 5.

# 2. Extrapolation

In this section we describe a straightforward technique to significantly increase the accuracy of a DDA simulation with a relatively small increase of computation time. This technique does not require any modification of a DDA program but only postprocessing of computed data. Therefore it can be easily implemented in any existing DDA code.

In Paper 1 we have proven that the error of any measured quantity is bounded by a quadratic function of the discretization parameter $y = kd|m|$ ($k$ – free space wave vector, $m$ – refractive index of the scatterer):

$$\left|\delta\phi^y\right| \leq \left(a_2^\phi - b_2^\phi \ln y\right)y^2 + \left(a_1^\phi - b_1^\phi \ln y\right)y, \qquad (1)$$

where $\phi^y$ is some measured quantity (e.g. extinction efficiency $Q_{\text{ext}}$, Mueller matrix elements at some scattering angle $S_{ij}(\theta)$, etc.) and $\delta\phi^y$ its error (difference between a result of the numerical simulation and an exact value). $a_{1,2}^\phi$, $b_{1,2}^\phi$ are constants (independent on $y$), which are described in detail in Paper 1.

Here we proceed and assume that for sufficiently small $y$, $\delta\phi^y$ can in fact be approximated by a quadratic function of $y$ (taking the logarithmic term as a constant). The applicability of this assumption will be tested empirically in Section 3.B. Introduction of higher-order terms is possible but not necessary (contrary to the quadratic term), and we avoid it in order to keep our technique as simple and robust as possible. We can now write:

$$\phi^y = a_0 + a_1 y + a_2 y^2 + \zeta^y, \qquad (2)$$

where $a_0$, $a_1$, $a_2$ are constants that are chosen such that $\zeta^y$ – the error of the approximation – is minimized. $a_0$ is then an estimate for the exact value of the measured quantity $\phi^0$. A procedure to determine $a_0$ is basically fitting of a quadratic function over several points $\{y, \phi^y\}$, which are obtained by a standard DDA simulation. In the ideal case of $\zeta^y = 0$ one can use any three values of $y$ to obtain the exact value of $\phi^0$. However, in practice different fits will always give different results. We limit ourselves to the usual least-square polynomial fit of the data. There are three question one should answer before conducting such a fit:

1) how many and which values of $y$ to use?
2) how to weight the influence of different calculated values used in the fitting, i.e. what is the behavior of expected errors $\zeta^y$? (Note that in the polynomial fit we minimize $\chi^2$, the summation of the squared difference between computed values and the fitting function weighted by the inverse of the expected error $\zeta^y$.)
3) how to estimate the difference between $a_0$ and $\phi^0$, i.e. the error of the final result?

It is important to note that, although there are some theoretical hints, answers to these questions are mainly empirical and should be tested. Our approach is based on the test cases presented in Section 3.B. These may not be representative for all scattering problems, but they do show the potential power of our approach. We do not attempt to choose the most suitable fit options, but merely demonstrate the applicability of the technique.

We start by analyzing the second question, i.e. what is the expected deviation from the quadratic model, i.e. what is the functional dependence of $\zeta^y$ on $y$, to be used as weighting function in the polynomial fitting procedure? For cubically shaped particles, defined in Paper 1 as particles whose shape can be exactly discretized using cubical subvolumes, one expects a smooth variation of the function $\phi^y$, and the error can be attributed as a model error, i.e. coming mainly from neglecting higher order terms in the convergence analysis of Paper 1. In that case the error $\zeta^y$ is expected to be a cubical function of $y$. We have tried cubical, quadratic and linear error functions when fitting results for cubically shaped particles and found that, although the differences are small, cubical errors generally lead to the best fits (data not shown).

Shape errors, which are present for non-cubically shaped particles, are expected to be very sensitive to $y$, because they depend upon the position of the particle surface inside the boundary dipole that changes considerably by a small variation of $y$ (for details see Paper 1).

Therefore shape errors can be viewed as random noise superimposed upon a smooth variation of $\phi^y$. The asymptotic behavior of shape errors is linear in $y$ (see Paper 1). Indeed, in certain cases we found that using linear errors $\zeta^y$ results in significantly better fits than when using cubical errors. However in other cases linear errors performed significantly worse. In our experience, using a cubical error function is in general always more reliable, even in the presence of shape errors, because it decreases the influence of points with high values of $y$, where the error is larger and less predictable. Since we want the procedure to be as robust as possible and not to use more complex error functions than strictly needed (e.g. polynomial), we take a cubical dependence of the error $\zeta^y$, both for cubically- and non-cubically shaped scatterers.

The choice of values of $y$ for computation can be described by the interval $[y_{min}, y_{max}]$, the number of points and their spacing. $y_{min}$ is usually determined by available computer hardware (time or memory bounds), that is the best discretization that can be computed for a given resource. The goal of the extrapolation procedure is to increase the accuracy beyond this "single DDA boundary". We will show in Section 3.B that the overall performance of this technique strongly depends upon $y_{min}$.

The choice of $y_{max}$ is governed by two notions: a larger interval of data points generally leads to better extrapolation but errors for high values of $y$ are more random and their significance is anyway much smaller (since we use a cubical error function). We have found that for cubically shaped scatterers a good choice is $y_{max} = 2 y_{min}$, while for non-cubically shaped scatterers increasing the interval to $y_{max} = 4 y_{min}$ does improve the fits. Probably that is due to the fact that the quality of fit for non-cubically shaped scatterers is determined by quasi-random shape errors and increasing the range leads to larger statistical significance of the result. We will also demand that $y_{max}$ is less than 1, since otherwise DDA is definitely far from its asymptotic behavior.

Spacing of the sample points depends partly on the problem, especially for cubically shaped scatterers (in that case an arbitrary number of dipoles cannot be used). We space computational points approximately uniform on a logarithmic scale, acknowledging the fact that a relative difference in $y$ is more significant than an absolute. The total number of points should be large enough for statistical significance. However, a large number of points increases computational time. We have used 5 points for cubically shaped particles (ratio of $1/y$ values is 8:7:6:5:4) and 9 points for non-cubically shaped particles (ratio of $1/y$ values is 16:14:12:10:8:7:6:5:4) or less if $y_{max} < 4 y_{min}$.

The estimation of the error of the final result is difficult since this error is due to model imperfection and not to some kind of random noise. The standard least-square fitting technique[22] provides a standard error (SE) for the parameter $a_0$, which we use as a starting point. Numerical simulations (Section 3.B) show that for spheres (the only non-cubical shape we studied) real errors are less than 2×SE in most cases. That is what one would expect if $\zeta^y$ is considered completely random (which is similar to the expected behavior of the shape errors). For cubical shapes, on the contrary, we have to estimate the error as 10×SE to reliably describe the real errors. It is important to note that an error estimate based on the SE is the simplest one can use. Its drawback is that we have to use a large multiplier (based on the real errors obtained in some of our simulations), which may lead to significant overestimation of real errors in certain cases.

We can now formulate the step-by-step extrapolation technique. We use abbreviations (c) and (nc) for cubically and non-cubically shaped scatterers respectively.
1) Select $y_{min}$ based on your computational resources.
2) Take $y_{max}$ to be 2 (c) or 4 (nc) times $y_{min}$ but not larger than 1.

3) Choose 5 (c) or 9 (nc) points over the interval [$y_{min}$,$y_{max}$] approximately uniformly spaced on a logarithmic scale.
4) Perform DDA computations for each $y$.
5) Fit the quadratic function (Eq. (2)) over the points $\{y,\phi^y\}$ using $y^3$ as errors of data points; $a_0$ is then the estimate of $\phi^0$.

Multiply SE of $a_0$ by 10 (c) or 2 (nc) to obtain an estimate of the extrapolation error.
Results of using this procedure are presented in Section 3, together with computational costs.

The extrapolation procedure is similar to a Romberg integration method,[22] which is adaptive. The error estimate, obtained by extrapolation, is an internal accuracy indicator of DDA computations that is just as important as the increase in the accuracy itself. Our error estimate opens the way to adaptive DDA, i.e. a code that will reach a required accuracy, using minimum computational resources.

## 3. Numerical simulations

### A. Discrete Dipole Approximation

The basics of the DDA method were summarized by Draine and Flatau.[2] In this paper we use the LDR prescription for dipole polarizability,[23] which is most widely used nowadays, e.g. in the publicly available code DDSCAT 6.1.[4] We also employ dipole size correction[6] for non-cubically shaped scatterers to ensure that the cubical approximation of the scatterer has the correct volume; this is believed to diminish shape errors, especially for small scatterers.[2] We use a standard discretization scheme without any improvements for boundary dipoles.

The main numerical challenge of DDA is to solve a large system of $3N$ linear equations. This is done iteratively using some Krylov-subspace method,[22] while the matrix-vector products are computed using an FFT-based algorithm.[5] Our code – Amsterdam DDA (ADDA) – is capable of running on a cluster of computers (parallelizing a *single* DDA computation), which allows us to use practically an unlimited number of dipoles, since we are not limited by the memory of a single computer.[24,25] We used a relative error of residual $<10^{-8}$ as a stopping criterion. Tests suggest that the relative error of the measured quantities due to the iterative solver is then $<10^{-7}$ (data not shown) and hence can be neglected (total relative errors in our simulations are $>10^{-6} \div 10^{-5}$ – see Section 3.B). All DDA simulations were carried out on the Dutch national compute cluster LISA.[26]

The execution time of one iteration depends solely on $N$, it consists of an arithmetic part which scales linearly with $N$ and an FFT part which scales as $N\ln N$. The number of iterations only slightly depends on the discretization parameter $y$ for fixed geometry of the scatterer. Rahola proved this theoretically for any Krylov-subspace method,[27] and our own experience agrees with this conclusion. Therefore the total computational time scales linearly with $N$ ($\propto y^{-3}$) or slightly faster (considering logarithm and imperfect optimization), which is consistent with our timing results (data not shown).

We can now estimate the computational overhead of the extrapolation technique compared to a single DDA computation for $y_{min}$ (time – $t(y_{min})$). Considering the spacing of points we used (described in Section 2) the execution time needed for 5 points computation is $t_5 < 2.5t(y_{min})$ and for the 9 points computation – $t_9 < 2.7t(y_{min})$. Memory requirements are the same as for a single computation. For comparison one should note that an 8 times increase in computational time and memory requirements (for single DDA computation with $y = y_{\min}/2$) gives only a 2 to 4 times increase in accuracy (depending in which error regime – linear or quadratic – $y_{\min}$ is located).

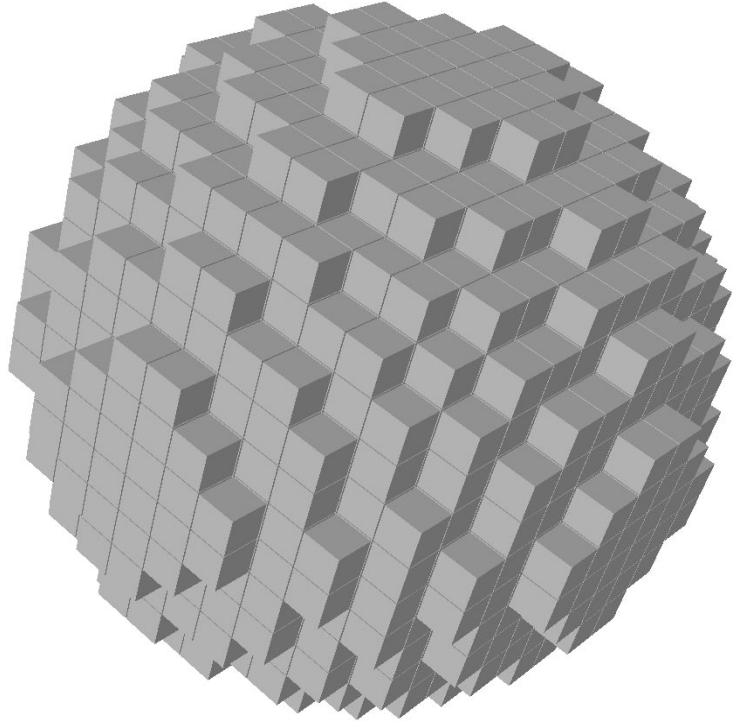

Fig. 1. Cubical discretization of a sphere using 16 dipoles per diameter (total 2176 dipoles).

## *B. Results*

We study five test cases: one cube with $kD = 8$, three spheres with $kD = 3, 10, 30$, and a particle obtained by a cubical discretization of the $kD = 10$ sphere using 16 dipoles per $D$ (total 2176 dipoles, see Fig. 1, $x$ equal to that of a sphere). By $D$ we denote the diameter of a sphere or the edge size of a cube. All scatterers are homogenous with $m = 1.5$. Although DDA errors significantly depend on $m$ (see e.g. [12]), we limit ourselves to one single value and study the effects of size and shape of the scatterer.

The maximum number of dipoles per $D$ ($n_D$) was 256. The values of $n_D$ that we used are of the form $\{4,5,6,7\} \cdot 2^p$ ($p$ is an integer), except for the discretized sphere, where all $n_D$ are multiples of 16 (this is required to exactly describe the shape of the particle composed from a number of cubes – see Fig. 1). The minimum values for $n_D$ were 8 for the $kD = 3$ sphere, 16 for the cube, the $kD = 10$ sphere, and the discretized sphere, and 40 for the $kD = 30$ sphere.

Typical computation time for the finest discretization (for the cube with $y = 0.047$, resulting in $N = 1.7 \cdot 10^7$) currently is 2.5 hours on a cluster of 64 P4-3.4 GHz processors. We expect that it can be improved by an order of magnitude by using modern FFT routines (e.g. fastest Fourier transform in the West – FFTW[28]) and a faster iterative solvers (bi-conjugate gradient stabilized or quasi-minimal residual that were shown to be clearly superior to CGNR[29,30] that we still use). We are currently improving our code along these lines.

All computations use a direction of incidence parallel to one of the principal axes of the cubical dipoles. The scattering plane is parallel to one of the faces of the cubical dipoles. In this paper we show results only for the extinction efficiency $Q_{\text{ext}}$ (for incident light polarized parallel to one of the principal axes of the cubical dipoles) and phase function $S_{11}(\theta)$ as the most commonly used in applications. However, the extrapolation technique is equally applicable to any measured quantity. For instance, we have also applied it for other Mueller matrix elements (data not shown).

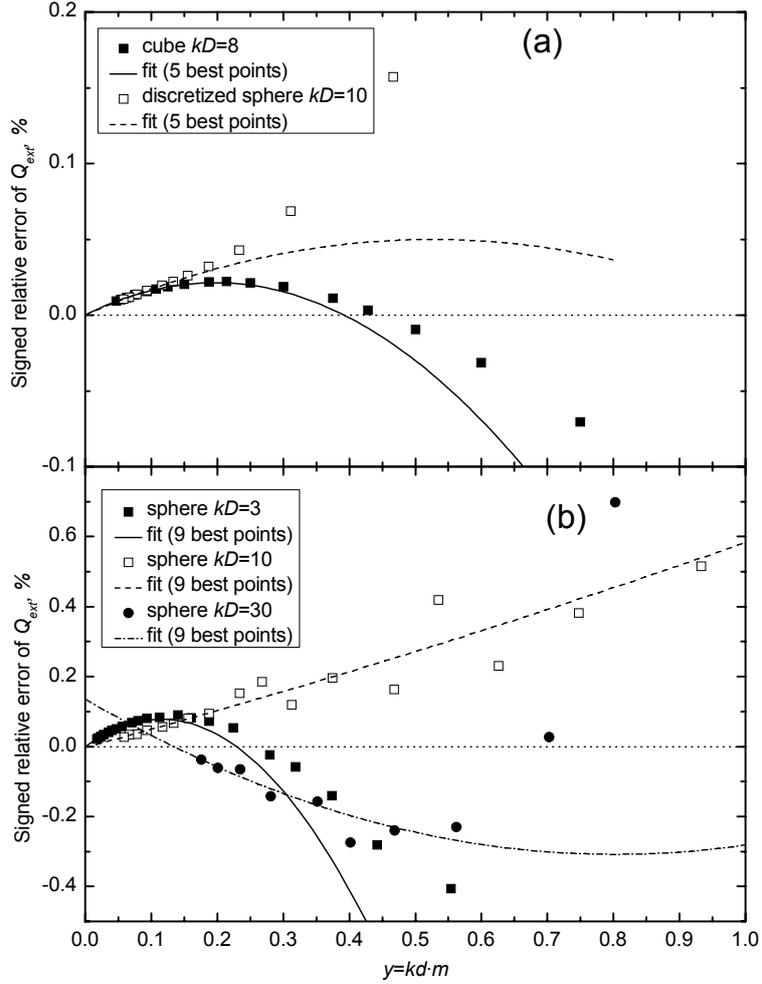

Fig. 2. Signed relative errors of $Q_{ext}$ versus $y$ and their fits by quadratic functions for (a) $kD = 8$ cube and discretized $kD = 10$ sphere, (b) 3 spheres. 5 and 9 best points are used for fits in (a) and (b) respectively.

Reference (exact) results of $S_{11}(\theta)$ and $Q_{ext}$ for spheres are obtained by Mie theory (the relative accuracy of the code we use[31] is at least $< 10^{-6}$). Unfortunately, no analytical theory is available for the cube and the discretized sphere, which could provide us with exact results. Instead, we use extrapolation over the 5 finest discretizations as reference results for these shapes.

To justify this choice we discuss, as an example, simulation results of $Q_{ext}$ for the cube. Instead of showing values of $Q_{ext}$ itself, we show in Fig. 2a $(Q_{ext}/a_0 - 1)$, with $a_0$ obtained through fitting the 5 finest discretizations. The extrapolation through these 5 best points ($y_{min} = 0.047$, $y_{max} = 0.094$) is also shown. The deviation of the fit from the five best points (that overlap on Fig. 2(a)) is very small indeed. This is also characterized by a small estimate of the extrapolation error $1.8 \times 10^{-6}$ (see Table 1). In Paper 1 we proved that DDA converges to the exact solution, therefore the result of the best extrapolation should be close to the exact result. The relative difference between the best discretization and the best extrapolation is only $9.0 \times 10^{-5}$, therefore it does not make a big difference which one to use as a reference when evaluating, for instance, the error of the extrapolation through the 5 worst discretizations ($y_{min} = 0.38$, $y_{max} = 0.75$). Hence all conclusions with respect to the reliability of the error estimates (as discussed in Section 4) do not depend on the choice of reference if $y_{min}$ is large enough. We also apply this reasoning to smaller $y_{min}$ and assume that using the reference value obtained by extrapolation of the finest discretizations is a good enough estimate of the exact value.

Table 1. Extrapolation errors of $Q_{ext}$. Estimate of the extrapolation errors is 10×SE for first two particles and 2×SE for spheres.

| $y_{min}$ | $y_{max}$ | Points | Error for $y_{min}$ | Extrapolation | |
|---|---|---|---|---|---|
| | | | | Estimate | Real |
| $kD = 8$ cube | | | | | |
| 0.047 | 0.094 | 5 | $9.0×10^{-5}$ | $1.8×10^{-6}$ | ——— |
| 0.094 | 0.19 | 5 | $1.6×10^{-4}$ | $6.6×10^{-6}$ | $4.6×10^{-6}$ |
| 0.19 | 0.38 | 5 | $2.2×10^{-4}$ | $5.3×10^{-5}$ | $4.0×10^{-5}$ |
| 0.38 | 0.75 | 5 | $1.1×10^{-4}$ | $3.7×10^{-4}$ | $3.2×10^{-4}$ |
| Discretized $kD = 10$ sphere | | | | | |
| 0.058 | 0.12 | 5 | $1.0×10^{-4}$ | $2.4×10^{-5}$ | ——— |
| 0.12 | 0.23 | 5 | $2.0×10^{-4}$ | $9.0×10^{-6}$ | $7.9×10^{-6}$ |
| 0.23 | 0.93 | 4 | $4.3×10^{-4}$ | $1.2×10^{-3}$ | $5.9×10^{-4}$ |
| $kD = 3$ sphere | | | | | |
| 0.018 | 0.070 | 9 | $2.2×10^{-4}$ | $1.0×10^{-5}$ | $4.1×10^{-6}$ |
| 0.035 | 0.14 | 9 | $4.0×10^{-4}$ | $5.9×10^{-5}$ | $4.8×10^{-5}$ |
| 0.070 | 0.28 | 9 | $6.8×10^{-4}$ | $8.7×10^{-5}$ | $5.7×10^{-6}$ |
| 0.14 | 0.54 | 9 | $9.0×10^{-4}$ | $3.7×10^{-4}$ | $7.0×10^{-4}$ |
| 0.28 | 0.54 | 5 | $2.4×10^{-4}$ | $4.3×10^{-3}$ | $1.8×10^{-3}$ |
| $kD = 10$ sphere | | | | | |
| 0.059 | 0.23 | 9 | $2.7×10^{-4}$ | $2.0×10^{-4}$ | $2.7×10^{-5}$ |
| 0.12 | 0.47 | 9 | $5.5×10^{-4}$ | $5.5×10^{-4}$ | $3.7×10^{-4}$ |
| 0.23 | 0.93 | 9 | $1.5×10^{-3}$ | $3.1×10^{-3}$ | $2.1×10^{-3}$ |
| $kD = 30$ sphere | | | | | |
| 0.18 | 0.70 | 9 | $3.8×10^{-4}$ | $1.3×10^{-3}$ | $1.4×10^{-3}$ |
| 0.18 | 0.35 | 5 | $3.8×10^{-4}$ | $3.3×10^{-3}$ | $6.9×10^{-4}$ |

Table 2. Comparison of shape and discretization errors of $Q_{ext}$ for $kD = 10$ sphere discretized with $y = 0.93$. All errors are relative to the best extrapolation result for the discretized sphere.

| | Shape | Discretization | Total |
|---|---|---|---|
| Error | $3.1×10^{-3}$ | $8.3×10^{-3}$ | $5.2×10^{-3}$ |

The same justification is valid for the discretized sphere (see Table 1 for $Q_{ext}$ results). Comparison of errors of different extrapolations results of $S_{11}(\theta)$ (shown in Fig. 3 and Fig. 4) is even more convincing. Reference results themselves (both of $Q_{ext}$ and $S_{11}(\theta)$) can be found in Paper 1.

Next we show the results obtained by the extrapolation technique. The dependence of the signed relative errors of $Q_{ext}$ on $y$ for all 5 test cases are shown in Fig. 2. Fig. 2(a) depicts results for the cube and the discretized sphere. The 5 best points for each scatterer are fitted by a quadratic function, using the method described in Section 2. Fig. 2(b) depicts extrapolation results for spheres, using the 9 best points for each of them (cf. Section 2). Since the exact Mie solution is available, intersection of a fit with a vertical axis is a measure of the accuracy of extrapolation result. Table 1 summarizes the parameters ($y_{min}$, $y_{max}$, number of points) of all the extrapolations, which were carried out, and their performance for $Q_{ext}$.

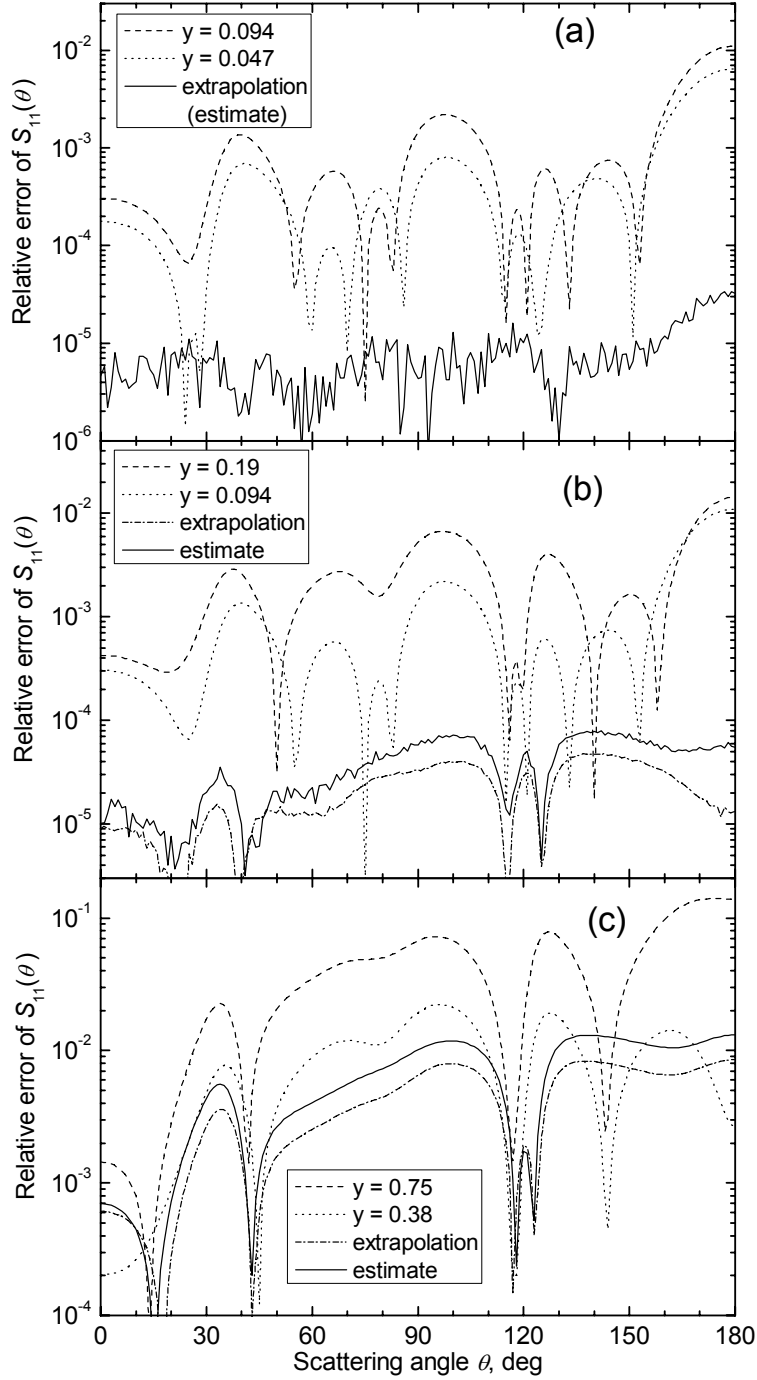

Fig. 3. Errors of $S_{11}(\theta)$ in logarithmic scale for extrapolation using 5 values of $y$ in the intervals (a) [0.047,0.094], (b) [0.094,0.19], and (c) [0.38,0.75] for $kD = 8$ cube. Estimate of the extrapolation error is 10×SE.

Next we present some of the extrapolations results for $S_{11}(\theta)$. Results for the cube are shown in Fig. 3. Each subfigure shows real (compared to the best extrapolation – reference) and estimated errors together with the errors of the finest and crudest discretizations used. Only the estimate of the error is shown for the best extrapolation – Fig. 3(a). Fig. 3(b) and (c) show extrapolation results using 5 points in the intervals [0.094,0.19] and [0.38,0.75] respectively. The performance of the extrapolation for the discretized sphere is shown in Fig. 4: (a) – best extrapolation, (b) and (c) – results for extrapolation using 5 and 4 points in the intervals [0.12,0.23] and [0.23,0.93] respectively. The broad spacing of points for extrapolation depicted in Fig. 4(c) is, as was noted above, due to the complex shape of the discretized sphere that limits possible values of $y$ to be 0.93 divided by an integer (total time for computing these 4 points is $< 1.6 t(y_{min})$). It is important to note once more that we use 10×SE as an estimate of extrapolation error for the cube and discretized sphere and 2×SE for spheres (cf. Section 2).

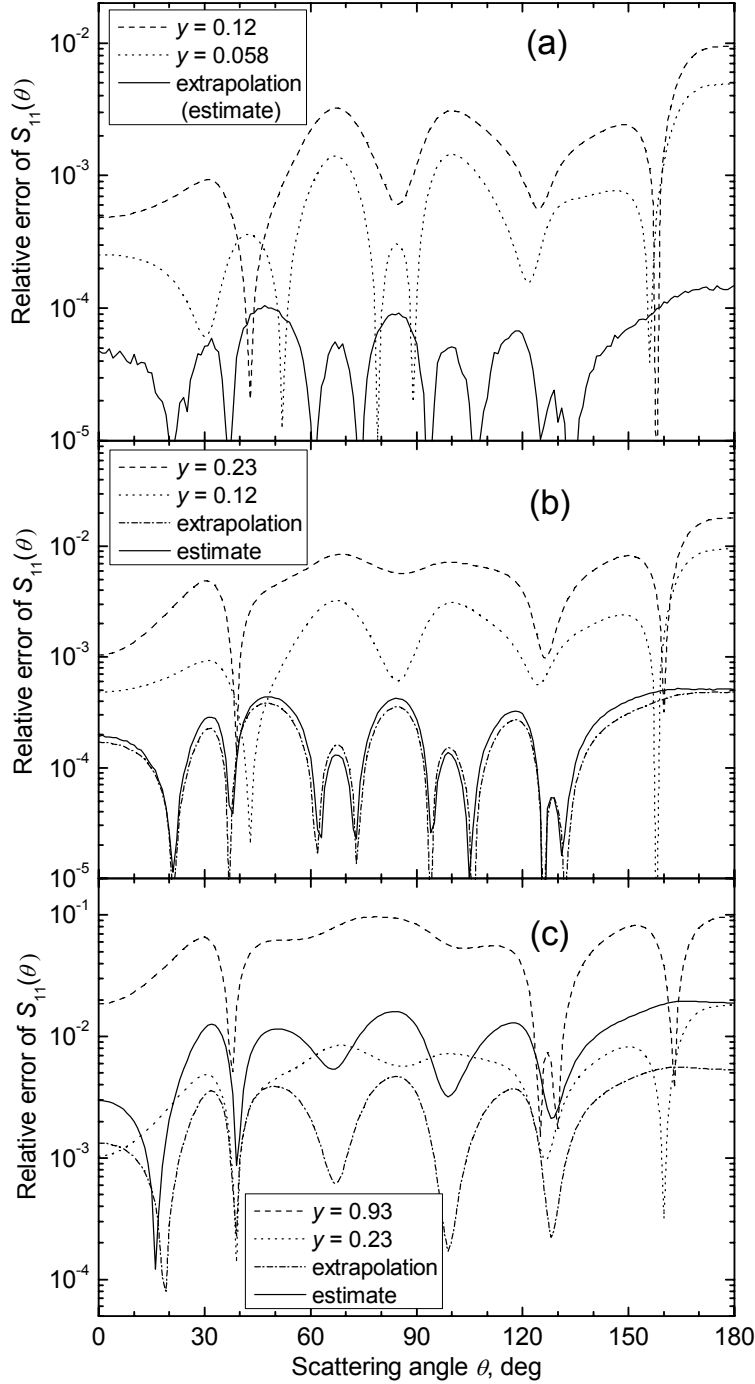

Fig. 4. Errors of $S_{11}(\theta)$ in logarithmic scale for extrapolation using 5 values of $y$ in the intervals (a) [0.058,0.12], (b) [0.12,0.23] ((c): 4 values of $y$ in the interval [0.23,0.93]) for the discretized $kD = 10$ sphere. Estimate of the extrapolation error is 10×SE.

Extrapolation results for the $kD = 3$ sphere are summarized in Fig. 5: (a) shows the best extrapolation (using 9 points in the interval [0.018,0.070]), and (b) shows the worst, but still satisfactory result, i.e. one that shows definite improvement of accuracy over most of the θ range. The extrapolation using 5 points from the interval [0.28,0.54] is no longer satisfactory (data not shown). Errors of the two best extrapolations for the $kD = 10$ sphere (using 9 points from the intervals [0.059,0.23] and [0.12,0.47]) are shown in Fig. 6(a) and (b) respectively. A third extrapolation for $kD = 10$ sphere is not satisfactory (data not shown). Both extrapolations for the $kD = 30$ sphere show similar controversial results, only one of them (9 points from the interval [0.18,0.70]) that is overall slightly better is shown in Fig. 7. The estimate of the extrapolation error is overall slightly higher than the real errors of the extrapolation (data not shown).

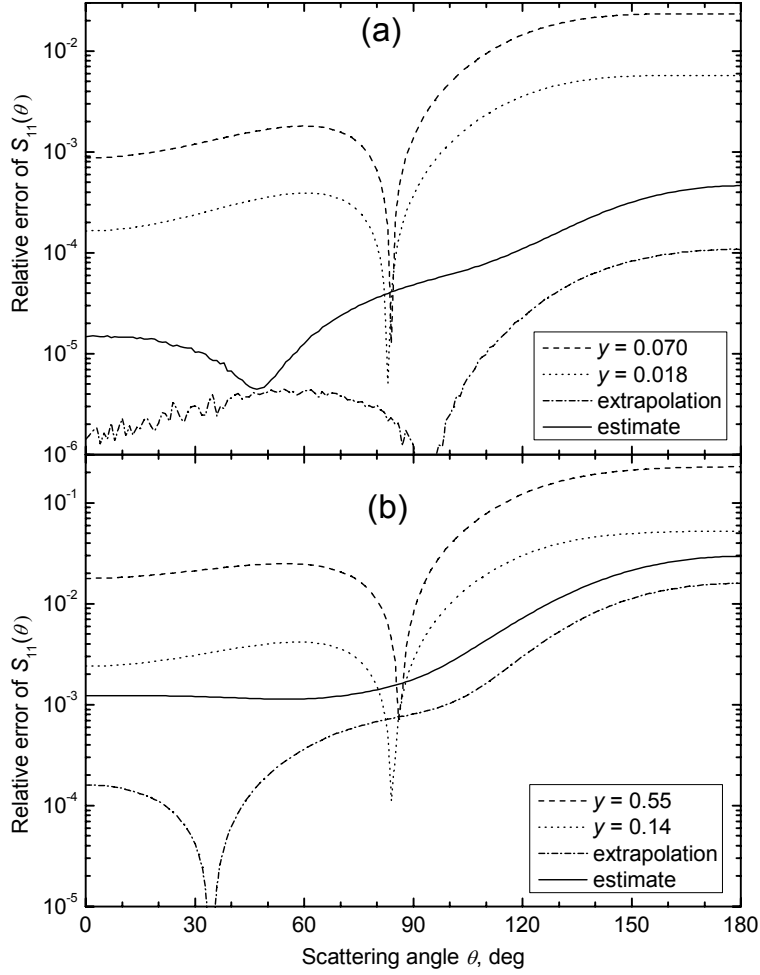

Fig. 5. Errors of $S_{11}(\theta)$ in logarithmic scale for extrapolation using 9 values of $y$ in the intervals (a) [0.018,0.070], (b) [0.14,0.55] for $kD = 3$ sphere. Estimate of the extrapolation error is 2×SE.

Results of $S_{11}(\theta)$ for all extrapolations (see Table 1) support the following trend: the quality of the extrapolation (defined as decrease of error compared to a single DDA computation for $y_{min}$) rapidly degrades with increasing $y_{min}$. The ratio of estimated to real errors increase with increasing $y_{min}$ (that can be considered as a degradation of the estimate quality).

Computation of exact results for both the $kD = 10$ sphere and its cubical discretization ($y = 0.93$) allows us for the first time to directly separate and compare shape and discretization error of single DDA computations. The shape error is the difference between some measured quantity for a discretized sphere (calculated to a high accuracy) and that for the exact sphere. The discretization error is difference between calculation using a limited number of dipoles (2176) and exact (very accurate) solution for the cubical discretization of the sphere (first curve in Fig. 4(c)). The total error is just the sum of the two. These three types of errors for $S_{11}(\theta)$ are shown in Fig. 8, all relative to the exact value for discretized sphere. Errors of $Q_{\text{ext}}$ are shown in Table 2.

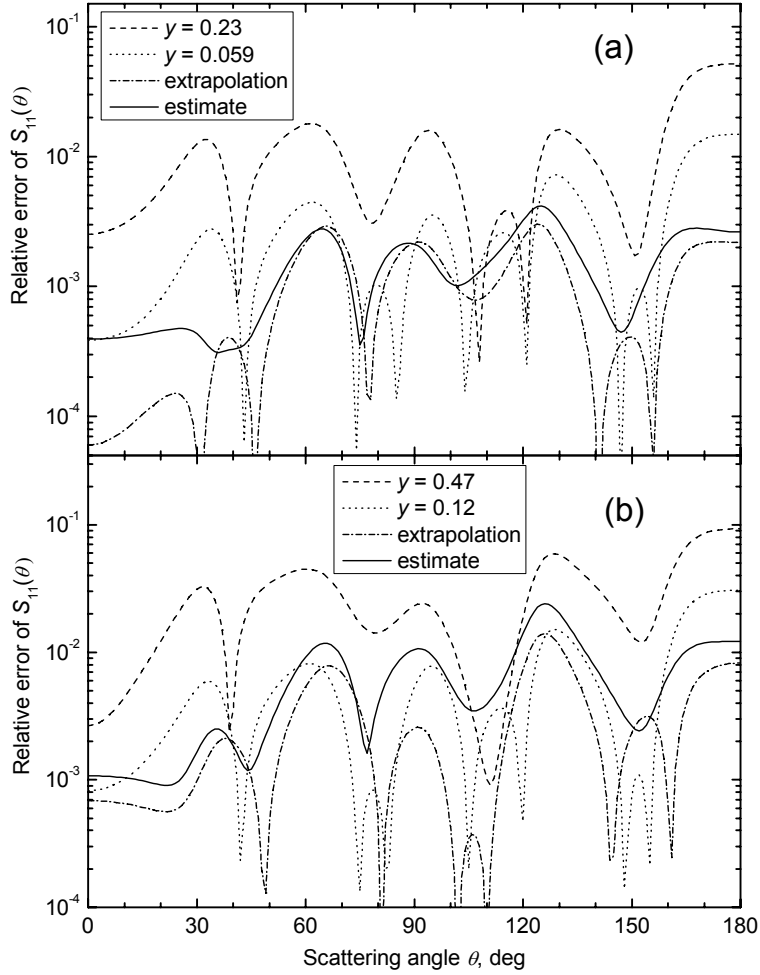

Fig. 6. Errors of $S_{11}(\theta)$ in logarithmic scale for extrapolation using 9 values of $y$ in the intervals (a) [0.059,0.23], (b) [0.12,0.47] for $kD = 10$ sphere. Estimate of the extrapolation error is 2×SE.

## 4. Discussion

In their review Draine and Flatau[2] gave the condition $y < 1$ for applicability of DDA. Usually $y = 0.6$ (10 dipoles per wavelength in the medium) is used in applications.[3] Smaller $y$ are used only in studies of DDA errors[2,12,13] or of light scattering by particles much smaller than a wavelength (then $d$ is determined by a shape of a scatterer, and $y$, being proportional to scatterer size, can be arbitrarily small).[32] However, if one wishes to achieve better (than usual) accuracy of a DDA simulation, smaller $y$ must be used.

The best extrapolation for the cube (Fig. 3(a)) shows a large improvement compared to the best single DDA calculation (it should be noted, however, that this result is based on the empiric error estimate). Maximum errors are decreased more than 2 orders of magnitude. This would be impossible to reach by a single DDA calculation as it will require over 6 orders of magnitude increase in execution time and memory, since there is only linear convergence for such small $y$. Even for $y_{\min} = 0.38$ the extrapolation can be called satisfactory because the maximum error is decreased almost two times when considering the estimate of the error (the real errors are even less). It is important to note that an estimate of the error is important by itself (even when it is not less than the error of a single DDA computation) because it does not require an exact solution (that is usually unavailable in real applications). In general, the extrapolation decreases large errors better than those that are already small, i.e. it may significantly decrease maximum errors but prove less satisfactory for certain measured quantity (e.g. $S_{11}$ for certain $\theta$). This conclusion holds true for all the extrapolations we performed (Fig. 3 – Fig. 7 and those not shown).

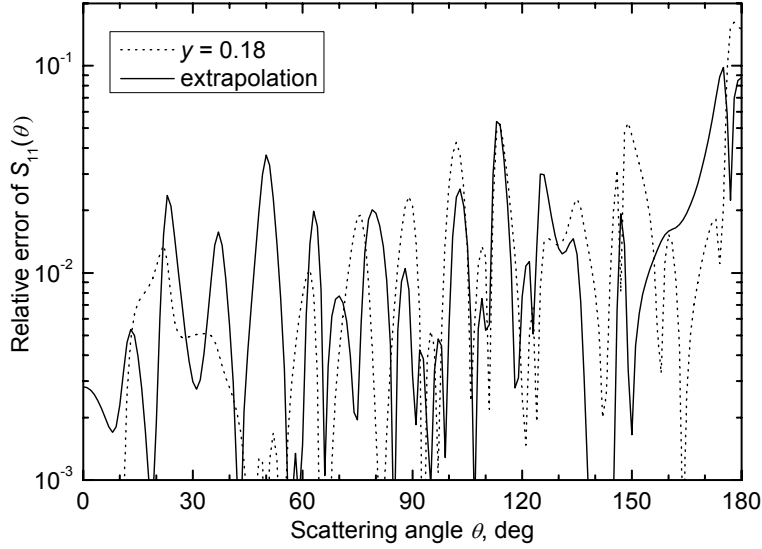

Fig. 7. Errors of $S_{11}(\theta)$ in logarithmic scale for extrapolation using 9 values of $y$ in the interval [0.18,0.70] for $kD = 30$ sphere.

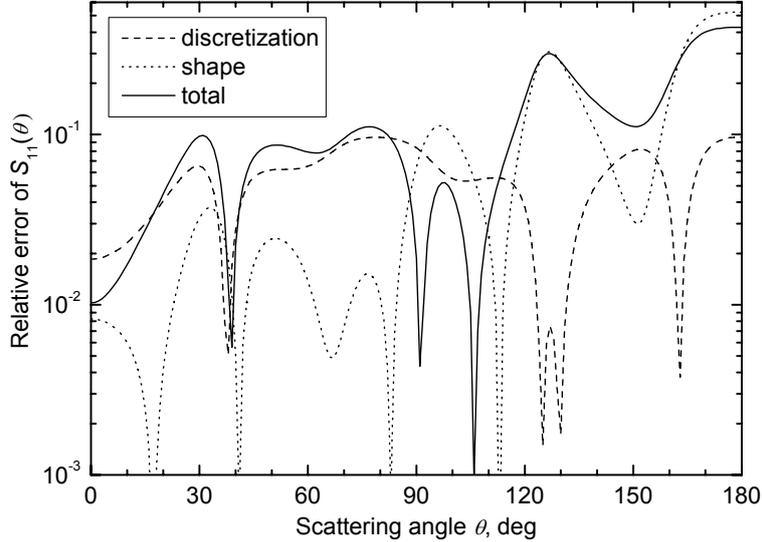

Fig. 8. Comparison of discretization and shape errors of $S_{11}(\theta)$ for $kD = 10$ sphere discretized using 16 dipoles per $D$ ($y = 0.93$).

Extrapolation results for the discretized sphere (Fig. 4) are similar to those for the cube. Extrapolations for $y_{min} = 0.058$ and 0.12 are very good (more than an order of magnitude decrease of maximum errors), while for $y_{min} = 0.23$ it is on the edge of being satisfactory. The latter is strongly influenced by the fact that only 4 points in a broad interval are used (hence it does not fully comply with the procedure specified in Section 2).

The best extrapolation for the $kD = 3$ sphere (Fig. 5(a)) shows results comparable to cubically shaped scatterers, however it uses an extremely small $y_{min} = 0.018$. Already for $y_{min} = 0.14$ (Fig. 5(b)) it only decreased the maximum errors by a factor of two. A similar boundary value of $y_{min}$ for satisfactory extrapolation is observed for $kD = 10$ sphere (Fig. 6(b)), while the best extrapolation (Fig. 6(a)) does show good results (4 times decrease of maximum error), although significantly worse than the analogous results for cubically shaped scatterers. Unfortunately we are currently not able to reach sufficiently small $y$ for the $kD = 30$ sphere and the best extrapolation (Fig. 7) uses rather large $y_{min} = 0.18$, resulting in almost negligible improvement of accuracy.

We have also studied a $kD = 8$ porous cube that was obtained by dividing a cube into 27 smaller cubes and then removing randomly 9 of them. All the conclusions are the same as those reported for the cube, but with slightly higher overall errors (data not shown).

Extrapolation of $Q_{ext}$ (Table 1) shows similar results as discussed above, however the improvement of accuracy is generally less than for maximum errors in $S_{11}(\theta)$ (which is in agreement with what we stated above, since errors in $Q_{ext}$ are already small). Moreover, one should take into account that errors of a single DDA calculation for some $y_{min}$ are unexpectedly small (e.g. the last extrapolations for the cube and the $kD = 3$ sphere), but these are just "lucky hits" near the points where the function $\delta Q_{ext}(y)$ crosses the horizontal axis (cf. Fig. 2).

Summarizing all results we can conclude that shape errors significantly degrade the extrapolation performance, because of its abrupt behavior, and therefore the extrapolation technique is much more suited for cubically shaped particles. One may expect satisfactory extrapolation for non-cubically shaped particles only when $y_{min} < 0.15$, while for cubically shaped particles the condition is $y_{min} < 0.4$. It is important to note though that extrapolation can be used for any $y_{min}$. The estimate of the error coming from the fitting procedure (SE) can then be used to decide whether this extrapolation was satisfactory or not. The quality of the extrapolation significantly increases with decreasing $y_{min}$, hence extrapolation is of biggest value for obtaining (very accurate) benchmark results. The size of the particle for which the extrapolation technique provides significant improvement is mainly determined by available computational resources that are required to reach small enough $y_{min}$. However, further testing is required to evaluate the quality of extrapolation for scatterers large compared to the wavelength.

It is important to note that the linear extrapolation that was applied in two papers[7,15] may lead to completely erroneous results (e.g. if points on the right branch of the parabolas for the cube and $kD = 3$ sphere in Fig. 2 are used). Quadratic extrapolation, as proposed in this paper, is much more reliable.

Throughout all the extrapolations we have used error estimates as specified in Section 2: 10×SE and 2×SE for cubically and non-cubically shaped scatterers respectively. All the results show that these estimates are reliable, i.e. in most cases real errors are less than the estimates. There are only two exceptions, both for the $kD = 3$ sphere: the fourth extrapolation of $Q_{ext}$ (Table 1) – real error 1.8 times larger than estimate – and second of $S_{11}$ – real error 1.5-2 times larger than estimate for broad range of $\theta$ (data not shown). The existence of such exceptions is acceptable since the estimates have a statistical nature of a confidence interval. However, these estimates, though reliable, are definitely not optimal, i.e. they often significantly overestimate the real errors (e.g. Fig. 5(a)). It also seems to be sensitive to the spacing of $y$ values used for extrapolation – cf. Fig. 4(c), where unusually broad spacing was used. Generally this overestimation increases with increasing $y_{min}$. We can conclude that the error estimate should be improved, and this is subject of future research. However, the current estimate is already suitable for practical applications since they mainly require reliability of the error estimate, which is demonstrated empirically in this paper.

It is important to note that we limited ourselves to a single value of *m*. While bounds of $y_{min}$ to obtain satisfactory extrapolation definitely dependent on *m*, other conclusions, such as the reliability of the error estimate, are expected to hold true for a broad range of $m$. This can be easily tested for specific values of $m$ of interest using the methodology put forward in this paper.

Finally we discuss the results presented in Fig. 8. One cannot conclude that shape errors dominate over discretization errors (or the other way around): for some $\theta$ shape errors are much larger than discretization, for others – vice versa. However, maximum errors occurring in backscattering directions are *definitely* due to shape errors (ratio of maximum shape to maximum discretization errors is about 4). Errors in $Q_{\text{ext}}$ (Table 2) are, on the contrary, mostly due to discretization (although they are almost two orders of magnitude smaller than maximum errors of $S_{11}$). One may expect shape errors to become even more important for smaller values of $y$, since the linear component of discretization errors is significantly smaller than that of shape errors (hence for large values of $y$ shape errors scale linearly and discretization – almost quadratically). Our single result principally shows different angle dependence of shape and discretization errors of $S_{11}$: shape errors have a clear tendency to significantly increase towards backscattering, while the general trend of discretization errors is uniform over the whole $\theta$ range.

We have presented a simple method to directly separate shape and discretization errors and only one result for illustration. All previous comparisons of shape and discretization errors had significant inherent interpretation problems that caused a lot of discussions about their conclusions.[16-20] Our method is free of such problems and therefore can be used for rigorous study of shape errors in DDA. For instance, it can help to directly evaluate the performance of different techniques to reduce such errors, e.g. weighted discretization (WD).[9] Discretization errors are then the limit one can achieve by drastically reducing shape errors.

We have used a traditional DDA formulation[2] to show that the extrapolation technique can be used with current DDA codes (e.g. DDSCAT[4]) without any modifications. However, as we showed in Paper 1 several modern improvements of DDA (namely integration of Green's tensor (IT)[33] and WD) should significantly change the convergence behavior of DDA computations and hence influence the performance of the extrapolation technique. IT should completely eliminate the linear term for cubically shaped scatterers. This will improve the accuracy especially for small $y$, and probably also improve the quality of the extrapolation for such scatterers. WD should significantly decrease shape and hence total errors for non-cubically shaped particles, moreover it should significantly decrease the amplitude of quasi-random error oscillations because it takes into account the location of the interface inside the boundary dipoles. Therefore WD should improve the quality of the extrapolation for non-cubically shaped scatterers. Testing of extrapolation performance of DDA using IT and WD is a subject of a future study.

## 5. Conclusion

Based on the theoretical convergence analysis as presented in Paper 1, we proposed an extrapolation technique together with a step-by-step prescription, which allows accuracy improvement of DDA computations. The performance of this technique was studied empirically and we showed that it significantly suppresses maximum errors of $S_{11}(\theta)$ when $y_{\min} < 0.4$ and $0.15$ for cubically and non-cubically shaped scatterers respectively (for $m = 1.5$). The quality of the extrapolation improves with decreasing $y_{\min}$ reaching extraordinary performance especially for cubically shaped particles – more than two order of magnitude decrease of error when $y_{\min} \approx 0.05$ for wavelength-sized scatterers with $m = 1.5$ (total computational time for extrapolation is less than 2.7 times that for a single DDA computation).

The proposed estimates of the extrapolation error were proven to be reliable, although they can be improved to decrease overestimation of the errors in some cases. This error estimate is completely internal, and hence can be used to create adaptive DDA – a code that will automatically refine discretization to reach a required accuracy.

We also proposed a simple method to directly separate shape and discretization errors. Maximum errors of $S_{11}(\theta)$ for the $kD=10$ sphere with $m=1.5$, discretized using 16 dipoles per diameter ($y=0.93$) are mostly due to shape errors, however the same is not true for all measured quantities. This method can be employed to rigorously study fundamental properties of these two types of errors and to directly evaluate the performance of different techniques aimed at reducing shape errors.

Our theory predicts that modern DDA improvements (namely IT and WD) should significantly change the performance of the extrapolation technique, however numerical testing of these predictions is left for future research.

## Acknowledgements

We thank Gorden Videen and Michiel Min for valuable comments on earlier version of this manuscript and Denis Shamonin for help with 3D graphics. Our research is supported by the NATO Science for Peace program through grant SfP 977976.